\documentclass[11pt]{article}
\usepackage{amsfonts}
\pdfoutput=1
\usepackage{latexsym,amsmath,amssymb}
\usepackage{graphics,epstopdf}
\usepackage[pdftex]{graphicx}
\numberwithin{equation}{section}
\begin{document}

\bigskip

\bigskip

\begin{center}
{\Large \textbf{Performance evaluation of different optimization techniques for coverage and connectivity control in backbone based wireless networks}}

\bigskip

\textbf{Khalid Khan} and \textbf{D.K. Lobiyal}

School of Computer and System Sciences, SC \& SS, J.N.U., New Delhi-110067., India%
\\[0pt]

 khalidga1517@gmail.com; dklobiyal@gmail.com \\[0pt]

\bigskip

\bigskip

\textbf{Abstract}
\end{center}

\parindent=8mm {\footnotesize {In this paper, performance evaluation of Newton-Raphson and Conjugate Gradient method has been studied in comparison to Steepest Decent method for coverage and connectivity control in backbone based wireless networks. In order to design such wireless networks, the main challenge is to ensure network requirements such as network coverage and connectivity. To optimize coverage and connectivity, backbone nodes will be repositioned by the use of mobility control based on above mentioned methods. Thus the network get self organized which autonomously achieve energy minimizing configuration. Furthermore by simulation using $MATLAB ~R2010a$, methods are compared on the basis of optimized cost, number of iterations and elapsed time i.e. total time taken to execute the algorithm.}}

\bigskip


\bigskip

\section {Introduction}

\begin{figure*}[htb!]
\begin{center}
\includegraphics[height=4cm, width=6cm]{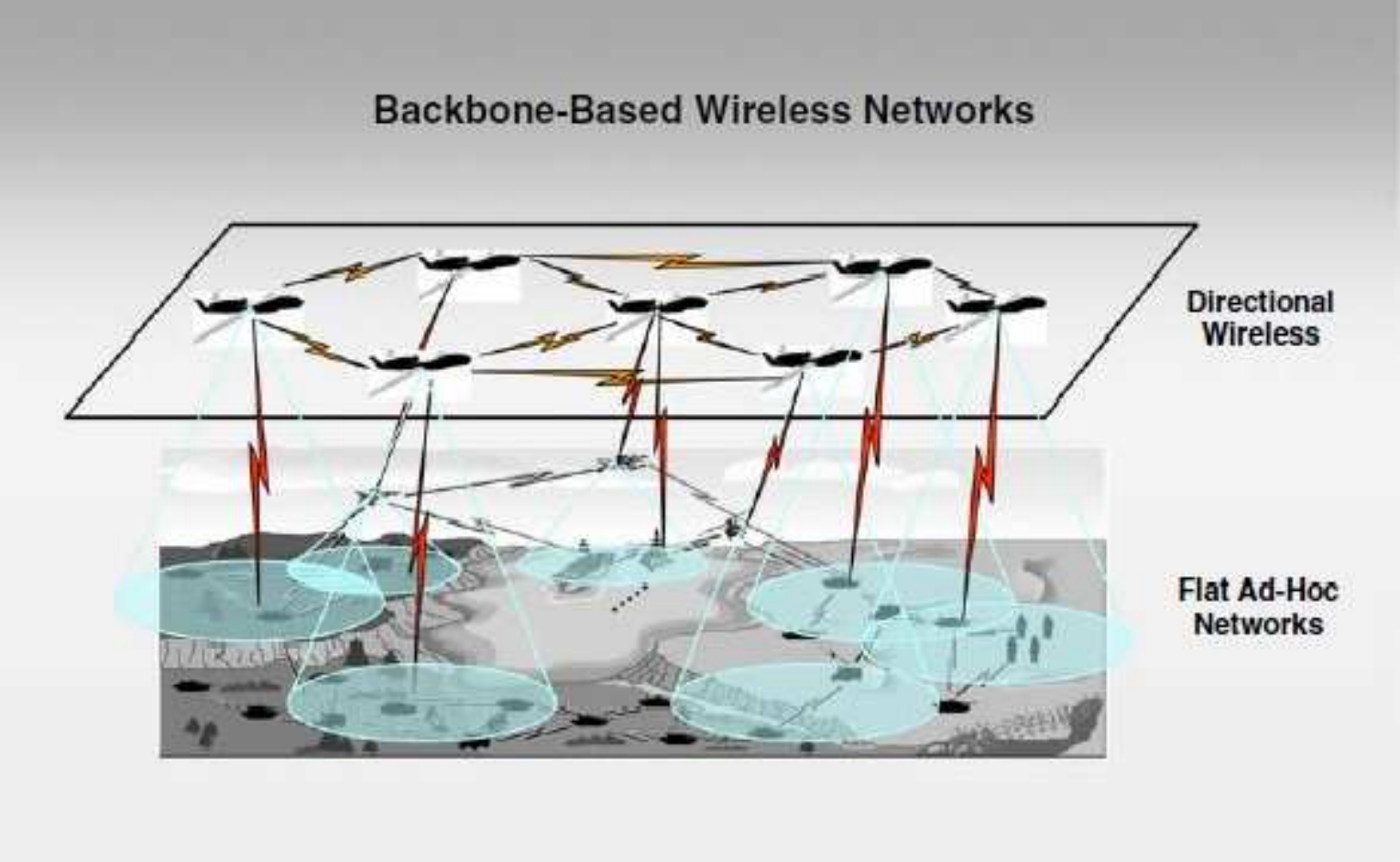}
\end{center}
\caption{Backbone based wireless network architecture \cite{lk}}\label{f16}
\end{figure*}

The need for assured end-to-end broadband connectivity and capacity limited flat ad hoc networks is driving communication networks to adopt layered wireless network architectures. These networks has diverse communication technologies and nodes with varying communication capabilities at different layers. In particular a two layer network architecture consists of nodes of higher capability at upper layer known as base stations or backbone nodes. These nodes use directional wireless communications such as Free Space Optical or directional Radio Frequency to provide broadband connectivity to capacity limited flat ad hoc networks known as host nodes at lower layer as shown in figure $\ref{f16}$. The use of directional wireless communications at upper layer can be exploited very well since high data rates in Gb/s are provided by point-to-point communication links and it also provides interference free communication \cite{dh,sd,sl}. These base station nodes form a backbone for the network since communication between two terminal host nodes takes place by multi-hop transmission scheme i.e. data sent by source (host node) first reaches to closest backbone node then it travels over the backbone network formed by backbone nodes until it reaches to that backbone node which is closest to the required destination node and from there it traverse few wireless nodes to reach the destination. \\

The most important concern in layered wireless network architecture is assurance of host nodes getting proper coverage and maintenance of connectivity among the backbone nodes in dynamic wireless networks. However providing coverage and maintaining connectivity both at a time is typically a competing objective. To maximize coverage, configuration of backbone nodes have to be done in a region where host nodes (wireless users) are deployed and to maintain robust backbone connectivity, these backbone nodes have to be brought close enough to each other. \\
Number of solutions has been proposed by various researchers related to the optimization problems based on location, deployment and tracking of sensor networks\cite{mkt,skm}. `Integer Linear Programming (ILP)' model has been given by Chakrabarty that minimizes the cost of heterogenous sensor nodes which guarantees coverage provided by sensor nodes \cite{ci}. Hinojosa et al. has formulated the problem of dynamic multi-product of facilities location in a mixed integer linear programming model and  minimized the total cost \cite{hpf}. Genetic algorithm is also applied by various authors to optimize the given network \cite{ak,nh}. The benefit of applying Genetic algorithm  is that function need not be continuous or differentiable. Similarly one of the solution is `Intersection point method' to overcome blind areas in deployment of sensors \cite{ab}. Details on wireless sensor networks can be found in \cite{ck,kd,mh,kn,sm,gk}.\\

Jaime Llorca et al.\cite{lk}  has optimized the joint coverage-connectivity problem  for quadratic cost function based on gradient approach. Here the actual energy usage in the network system is based on the quadratic cost functions for both coverage and connectivity which has been defined in terms of the square of the distances between neighboring nodes. The net force on a backbone node is defined as the energy gradient at the location of the backbone node, and only depends on neighbor's position information. Further Haijun Zhang et al. used Flocking algorithm and Particle Swarm Optimizer for similar joint coverage-connectivity optimization problem in \cite{haijun}.\\

In this paper, we study performance evaluation of Newton-Raphson and Conjugate Gradient method in comparison to Steepest Decent method for the quadratic minimization problem formulated by Jaime Llorca et al. in \cite{lk}.

\section{Problem Formulation}
Let us recall the quadratic minimization problem  given by Jaime Llorca et al. in \cite{lk}. Assume that there is a network of $M$ backbone nodes and $N$ host nodes located in a region $A$, subset of $R^2$ in the plane. The host nodes are located at $h_1, h_2,...,  h_N$ in the network, where $h_i = (x_{i}, y_{i})$ represents the location of $i^{th}$ host node in $ A$. Similarly, the backbone nodes are located at $B_1, B_2,..., B_M$ in the network, where $ B_j = (X_j, Y_j)$ represents the location of $j^{th}$ backbone node in $A$. Coverage is provided to each host nodes which are in close proximity of backbone nodes. The process in which host node $s$ will communicate with another host node $d$ will be done in the following way: firstly host node $s$ will transmit its data to the closest backbone node, from where it will traverse the backbone network until it reaches to that backbone node which is closest to the destination node. Finally, the backbone node transmit the data to intended host node $d$.\\

Therefore this scheme depends on two aspects, firstly the backbone nodes must provide proper coverage to host nodes, which means that each host node is in the close proximity of at least one backbone node and secondly there must be good connectivity between backbone nodes.\\

Both of these aspects are formulated as Coverage cost and Connectivity cost in \cite{lk}.\\

\noindent \textbf{Coverage cost}

Coverage is defined as how well a given region can be monitored. The Coverage cost is defined as the function of square distance of each individual host node to that backbone node which is providing coverage to it. If we increase the distance of host nodes from backbone nodes then coverage will decrease and after some more increment in distance host nodes may be in out of coverage mode and thus this coverage cost will increase on moving the backbone nodes far away from the host nodes. Assuming that $bi(j)$ represents the index of the backbone node that provides coverage to the host node $J$, where  $1\leq bi(j) \leq M$. Now we can define coverage cost as:
\begin{equation}\label{e21}
f_{1}=\sum\limits_{j=1}^{N}{\vert h_{j}-B_{bi(j)}\vert}^{2}
\end{equation}

where $f_1$ represents the coverage cost. This cost function will represents the energy usage by the backbone nodes in providing coverage to host nodes.\\

\noindent \textbf{Connectivity cost}

For maintaining connectivity between backbone nodes they can be simply connected to form a chain topology in which a single link connects all backbone node. But these types of simple graph are vulnerable to link failures as after any single link fail, the network becomes disconnected. Therefore it is very important to tolerate fault at some degree by maintaining proper connectivity in network. For a given connected network, the Connectivity cost is defined as the distance between connected backbone nodes. The connectivity cost function will increase on moving the backbone nodes far away from each other.\\

Connectivity cost is defined as:
\begin{equation}\label{e22}
f_{2}=\sum\limits_{j=1}^{M}\sum\limits_{k=1}^{M}{\vert B_{j}-B_{k}\vert}^{2}C_{jk}=tr(\Lambda C)
\end{equation}

where $f_2$ represents the connectivity cost, $tr(.)$ denotes the trace of a square matrix, and $C $ is an $M \times M$ matrix that represents the backbone connectivity graph, i.e., $ C_{jk} = 1$ if there exists a link between backbone nodes $j$ and $k$, moreover $\Lambda_{jk}= {\vert B_{j}-B_{k}\vert}^{2} $ and therefore $\Lambda$ is the $M \times M$ distance squared matrix of the backbone nodes with $\Lambda_{jk}$ as its $jk$ entry. This cost function will represents the energy usage by network to keep the backbone nodes connected.\\

Now the total cost function  can be defined as the sum of coverage cost and connectivity cost :
\begin{equation}\label{e23}
f=f_{1}+\lambda f_{2}=\sum\limits_{j=1}^{N}{\vert h_{j}-B_{bi(j)}\vert}^{2}+
\lambda tr(\Lambda C)
\end{equation}

which gives the total energy usage for the given network. In the above cost function relative importance of both the aspect i.e. coverage and connectivity is maintained by positive coefficient $\lambda$. For smaller value of $\lambda$, coverage is more important, while for large value of $\lambda$, higher priority will be given to connectivity and tries to keep the backbone nodes close to each other.
\section{Optimization Algorithms}


To optimize the objective function, certain iterative schemes has been employed for which search direction will be computed first and then decide, how far to move along that direction. The success of this iteration depends on effective choice of both direction $d_k$ and step length $\delta_k.$
 The algorithms Steepest descent and Newton Raphson which has been used in this paper require direction $d_k$ to be a descent direction, one for which $\langle \nabla f(x^{(k)}),d_k \rangle <0,$ as this property guarantees that the function $f$ can be reduced along this direction ( where ${x^{(k)}}$ denotes $x \in {R^{2}}$ at $k^{th}$ iteration) i.e.
 \begin{equation*}
  \text{If}~ \langle \nabla f(x^{(k)}),d_k \rangle <0,
 \end{equation*}
 \begin{equation*}
  \text{then}~ f(x^{(k)}+\delta_k d_k)~< f(x^{(k)}),~ \forall~ \delta_k \in(0, \delta_0).
 \end{equation*}
 where $\langle , \rangle $ is the notation for {\it usual inner product} between vectors in $R^2.$

\subsection{Steepest Descent Method}
In Steepest descent method, optimization of cost function (i.e. providing proper coverage and maintaining connectivity) is governed by the iterative scheme given as follows
\begin{equation}\label{e28}
 x_{j}^{(k+1)}=x_{j}^{(k)}+\delta_k~d_{k}
\end{equation}
where $x_{j}^{(k)}$   denotes the location of the backbone node $j$ at the $k^{th}$ iteration,  $\delta_k$ is a step size and $d_{k}$  is steepest descent direction.\\

%
%
 The steepest descent direction at the location of the $j^{th}$ backbone node can be defined as:
\begin{equation}\label{e27}
d_{k}=-\nabla f_{j}
\end{equation}
where
$\nabla f_{j}$ denotes the gradient of function  $f$ at $j^{th} $ backbone node.

To compute step length $\delta$, we know that a quadratic function $f(x)$, which can be expressed as
 \begin{equation*}
   f(x) = \frac{1}{2}\langle x,Qx\rangle - \langle b,x\rangle
 \end{equation*}where Q is symmetric and positive definite matrix and $x \in R^2.$\\

   \begin{equation*}
 \text{Let}~~~~~\phi(\delta) = f(x^{(k)} + \delta d_k) = f(x^{(k)} - \delta \nabla f(x^{(k)}))
\end{equation*}
 \begin{equation*}
 \Rightarrow \phi(\delta) = \frac{1}{2}\langle x^{(k)} - \delta \nabla f(x^{(k)}), Q (x^{(k)} - \delta \nabla f(x^{(k)}))\rangle - \langle b , x^{(k)} - \delta \nabla f(x^{(k)})\rangle
\end{equation*}
 For suficiently small value of $\delta$, we have $\phi^{'}(\delta) = 0$\newline
 \begin{equation}\label{e31}
 \Rightarrow \delta_k = \frac{\langle \nabla f(x^{(k)}),\nabla f(x^{(k)})\rangle}{\langle \nabla f(x^{(k)}) , Q \nabla f(x^{(k)})\rangle}
\end{equation}

so the relocation of backbone nodes in Steepest descent method will be governed by equation (\ref{e28}) based on step size given in (\ref{e31}).\\

Since in Steepest descent method, the direction of descent at every iteration is perpendicular to descent direction at previous iteration which makes this process slow as it follows zig-zag path to reach the optimized solution. Hence cost function has also been optimized by using Newton-Raphson and Conjugate Gradient method and it will be verified that the Conjugate Gradient method will take less time to optimize the cost function but there also exists tradeoff among these algorithms which has been shown later in simulation part. For performance evaluation, the results given by these algorithms has been compared based on optimized cost, number of iterations and elapsed time i.e. total time taken to execute the algorithm.

\subsection{Newton-Raphson method }

In general for the real valued function $f(x_{1},x_2),$ if all second order partial derivatives of $f$ exist and are continuous over the domain of the function, then the Hessian matrix of $f$ is given by
\begin{equation}\label{e41}
H= {\nabla}^{2} {f(x)}=\left(\begin{array}{cccccc}
                              \frac{ \partial^{2}f}{\partial x_{1}^2} & \frac{ \partial^{2}f}{\partial x_{1}\partial x_{2}}  \\
                               \frac{ \partial^{2}f}{\partial x_{2}\partial x_{1}} &  \frac{ \partial^{2}f}{\partial x_{2}^2}  \\

                             \end{array}
                           \right)
\end{equation}
\newline
where $x = (x_1, x_2).$\newline

\noindent A matrix is said to be positive definite if all its eigen values are positive.\\
For function  $f(x)$ which is strongly convex and twice differentiable,
 the iterative sequence using Newton's method \cite{aw} will be
 \begin{equation}\label{e28}
 x_{j}^{(k+1)}=x_{j}^{(k)}+\delta_k~d_{k}
\end{equation}

where descent direction $d_{k}$ is given by
\begin{equation*}
    d_k = -(\nabla ^2 f(x^{(k)}))^{-1} \nabla f(x^{(k)})
\end{equation*}
and $\delta_k=1$ has taken. Thus by Newton's method, iterative scheme is given as
\begin{equation*}
   x^{(k+1)} = x^{(k)} - (\nabla ^2 f(x^{(k)}))^{-1} \nabla f(x^{(k)})
\end{equation*}

\subsection{Conjugate-Gradient Method}
In this method, in place of descent direction, set of conjugate vectors are generated which are known as conjugate direction along which optimization of cost function will take place. First conjugate direction $d_0$ will be steepest descent direction at the initial point $x_0$. After this successive one-dimensional minimizations will take place along each of the conjugate directions $d_k$ until convergence is achieved. For details one can refer \cite{aw,nw}.\\



The iterative scheme is given by $x^{(k+1)} ~=~ x^{(k)} ~+~ \delta_k ~d_k,$ 
where
\begin{equation*}
\delta_k = \frac{- \langle {gr}_k,d_k \rangle}{\langle d_k,H_k d_k \rangle}.
\end{equation*}
 and Conjugate direction is governed by equation\\
$d_{k+1}~ =~ -{gr}_{k+1}~ +~ \beta_k~d_k$\\

where
\begin{equation*}
\beta_k = \frac{\langle {gr}_{k+1},H_k d_k \rangle}{\langle d_k,H_k d_k \rangle}
\end{equation*} and ${gr}_k$ denotes gradient at $k^{th}$ iteration.

\newpage

\section{Simulation Results}

   In order to compare the performance of various algorithms used for coverage and connectivity control in adhoc network, the optimization algorithms are implemented using $MATLAB ~R2010a$. For simulation purposes, a network of $20$ host nodes, uniformly distributed in a $100 \times 100$ two dimensional plane has been considered. Initially the first host node is placed randomly any where in the plane and other host nodes are placed within cluster range of the previous host node with probability $p$ and uniformly in the $100 \times 100$ plane with probability $(1 - p)$.  After placement of host nodes, $10$ backbone nodes are placed at random in the plane and an initial ring topology is built connecting all backbone nodes. Then the Optimization algorithm is executed to make backbone nodes adjust their position until convergence to the optimized network is achieved. The performance of different optimization algorithms namely Steepest Descent, Newton Raphson and Conjugate Gradient method are evaluated in terms of optimized cost, number of iterations and execution times. All results are compared on the basis of average over $20$ independent networks.\\

\begin{figure}[htb!]
\begin{center}
\includegraphics[height=4cm, width=6cm]{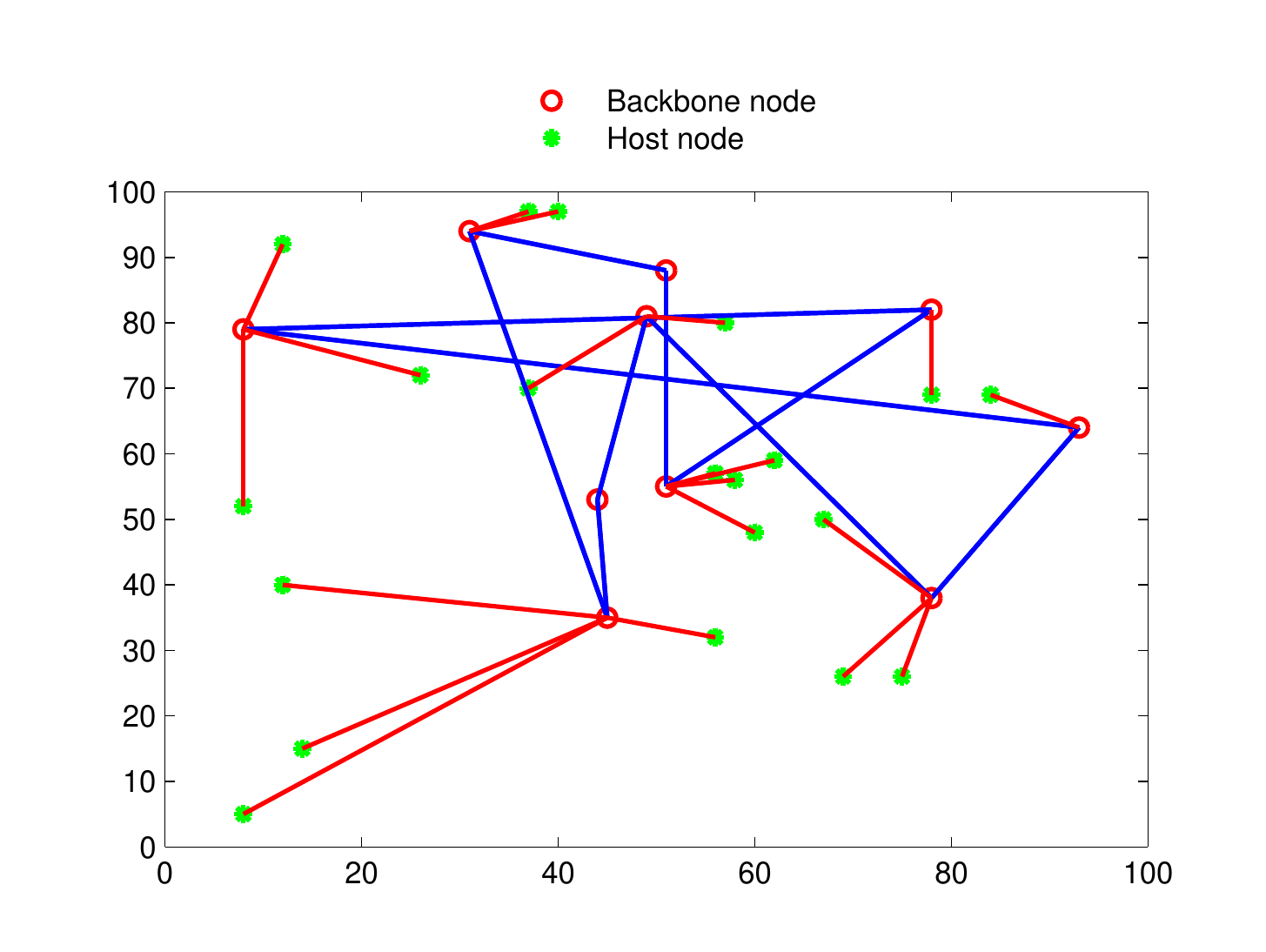}
\end{center}
\caption{Initial Network Configuration}\label{f1}
\end{figure}

 Figure $\ref{f1}$ shows the randomly generated initial network of $20$ host nodes and $10$ backbone nodes. The backbone nodes are shown with red circle and links between them are shown with blue lines. Each host nodes are shown with green circle and links between backbone nodes and host nodes are shown with red lines.\\

\begin{figure}[htb!]
\begin{center}
\includegraphics[height=4cm, width=6cm]{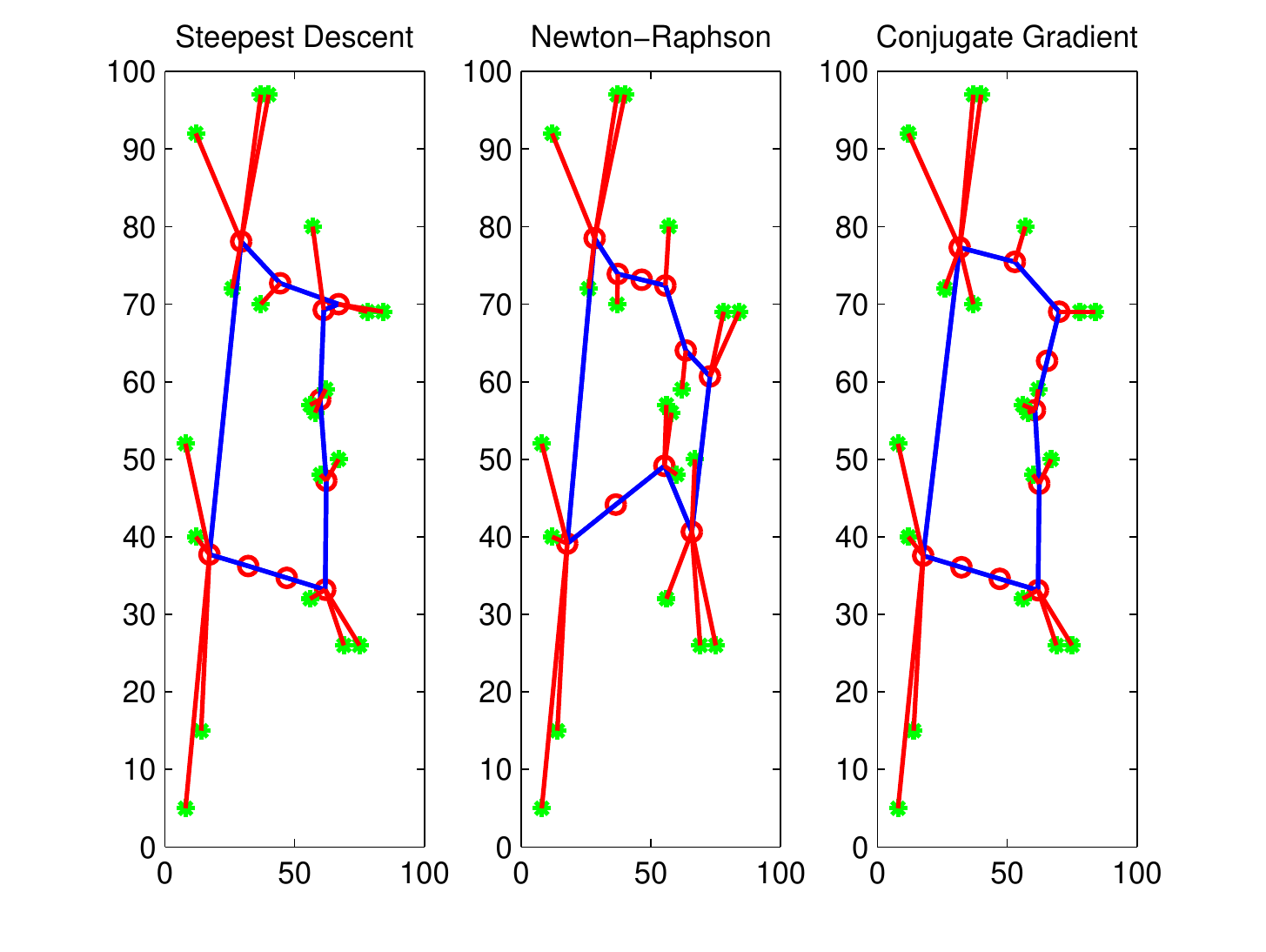}
\end{center}
\caption{Optimized network by different algorithms for $ \lambda =1$}\label{f2}
\end{figure}

\begin{figure*}[htb!]
\begin{center}
\includegraphics[height=4cm, width=6cm]{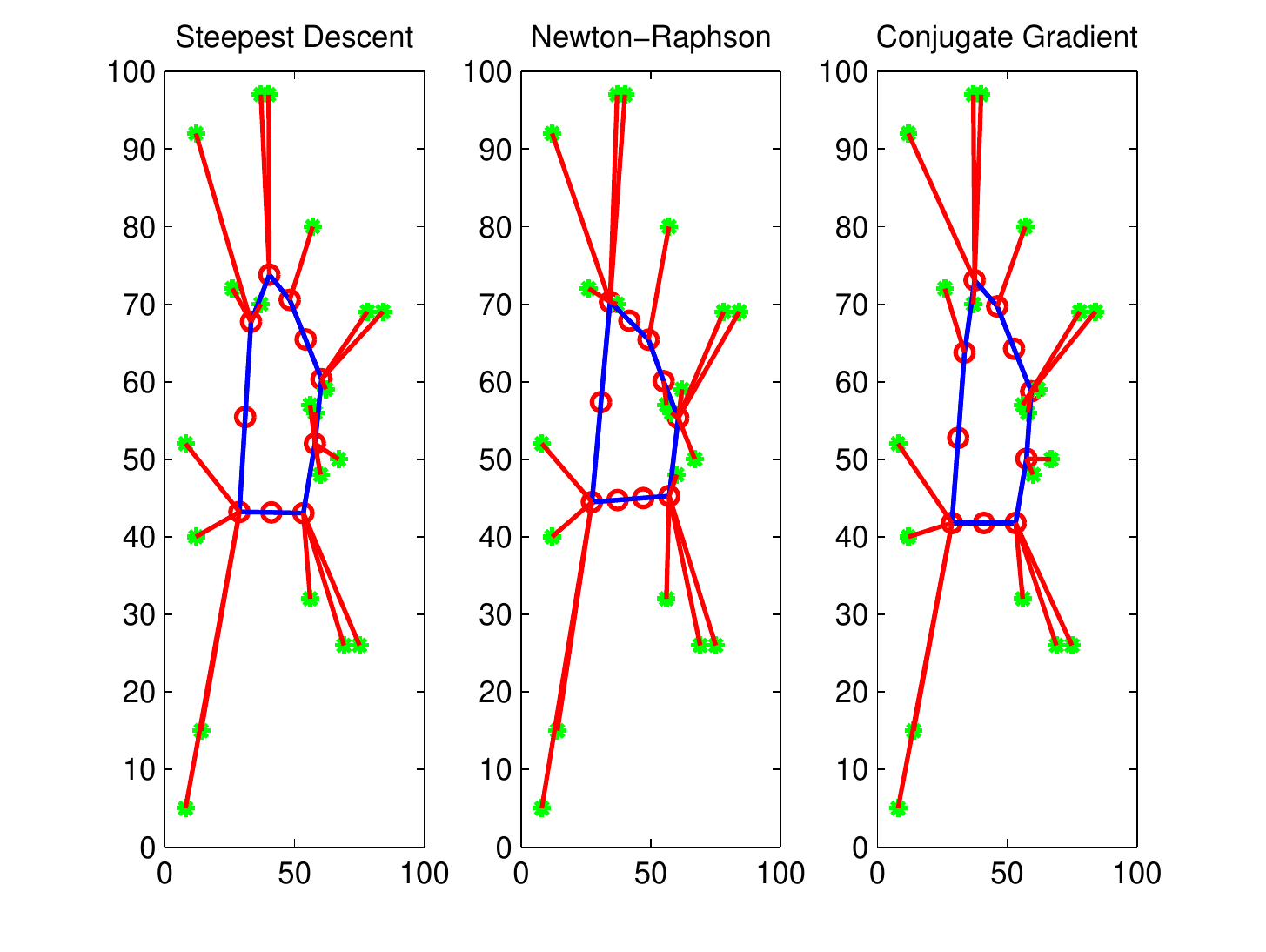}
\end{center}
\caption{Optimized network by different algorithms for $ \lambda =5$}\label{f3}
\end{figure*}

\begin{figure*}[htb!]
\begin{center}
\includegraphics[height=4cm, width=6cm]{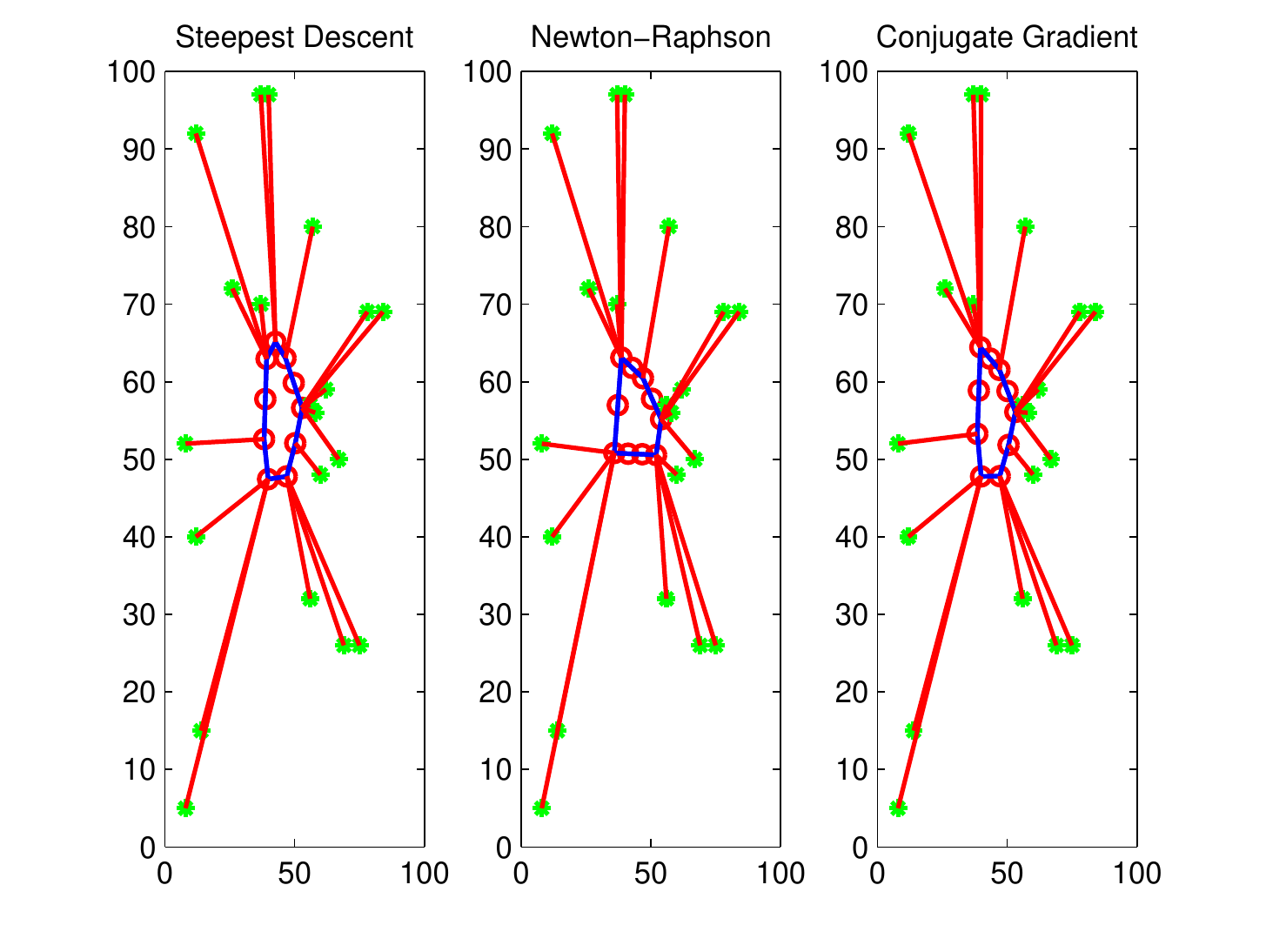}
\end{center}
\caption{Optimized network by different algorithms for $ \lambda =15 $}\label{f4}
\end{figure*}

\begin{figure*}[htb!]
\begin{center}
\includegraphics[height=4cm, width=6cm]{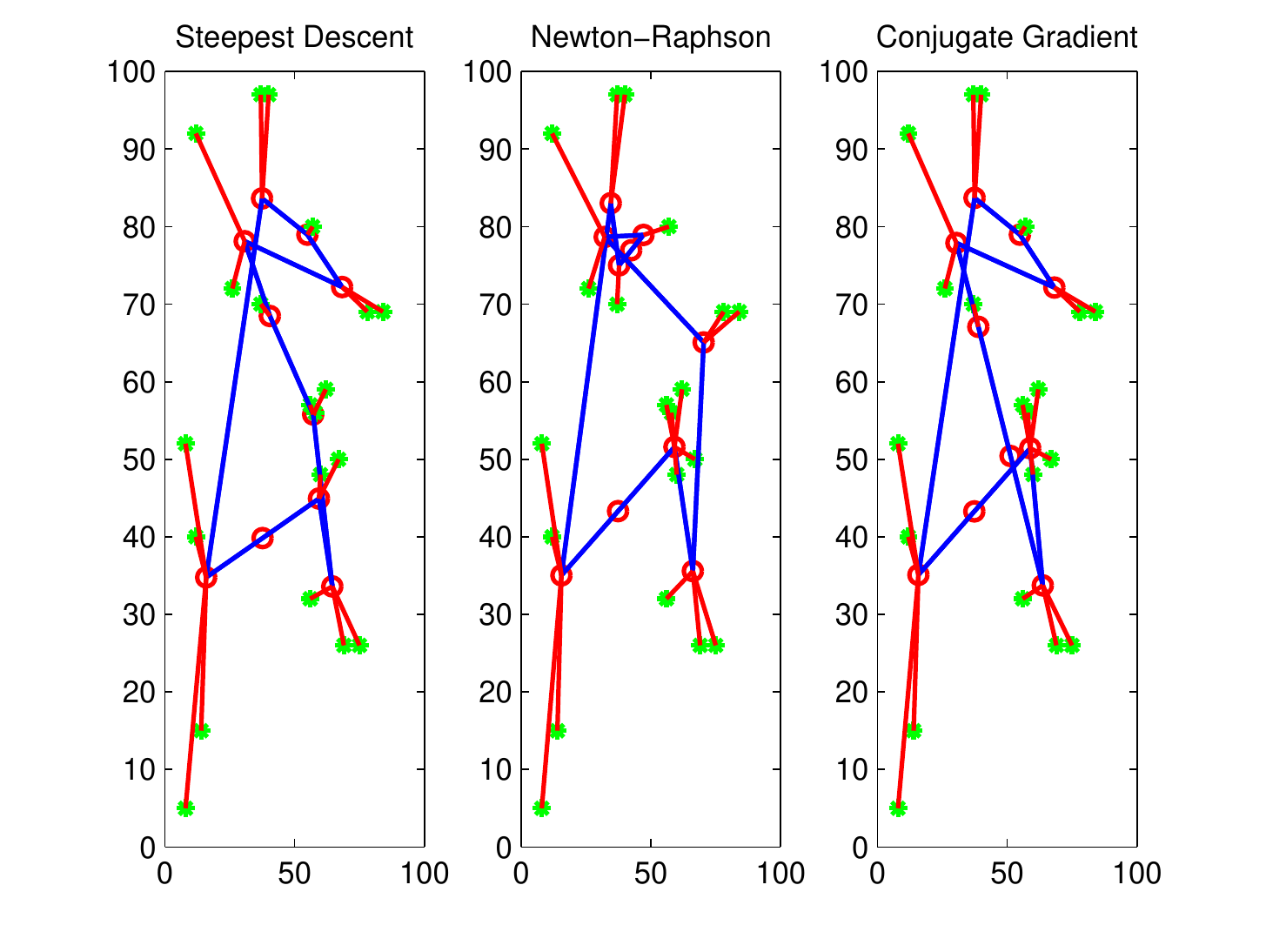}
\end{center}
\caption{Optimized network by different algorithms for $ \lambda =.5$}\label{f5}
\end{figure*}

\begin{figure*}[htb!]
\begin{center}
\includegraphics[height=4cm, width=6cm]{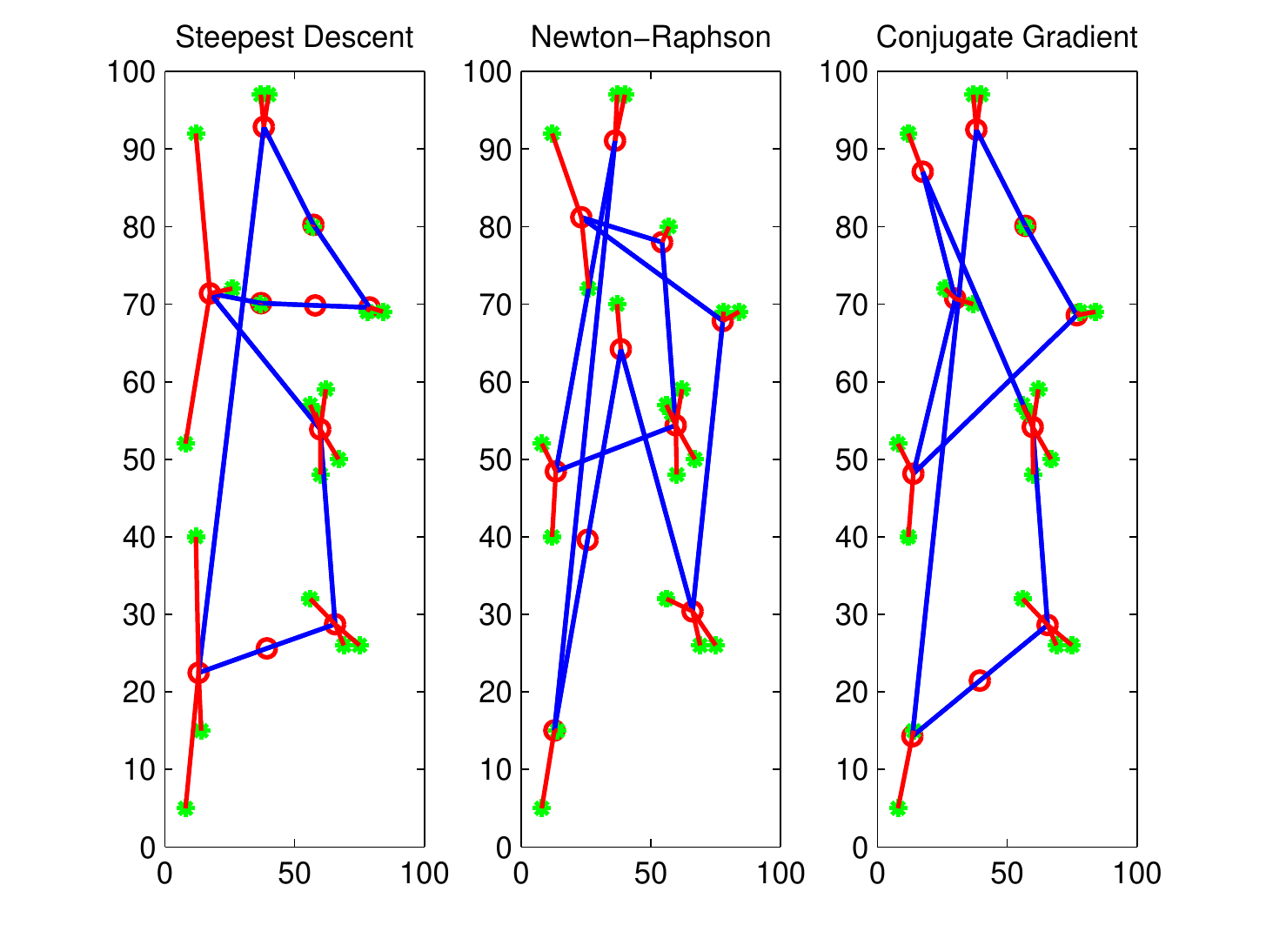}
\end{center}
\caption{Optimized network by different algorithms for $\lambda =.1$}\label{f6}
\end{figure*}

For the initial network given in figure $\ref{f1}$, the optimized network obtained using different algorithms is shown in figures $\ref{f2}$ for $\lambda=1.$ From equation \ref{e23}, it is clear that when $\lambda=1$, coverage and connectivity has been assigned equal weightage. Further as the value of $\lambda$ increases, weightage of connectivity increases at the cost of coverage. In order to verify the effect of increasing $\lambda$, the optimized network for $\lambda =5 ~\text{and}~15$ are shown in figures $\ref{f3}~\text{and}~\ref{f4}$ respectively. It can be seen from these figures that backbone nodes start coming closer to each other in order to maintain more connectivity. Similarly, when value of $\lambda$ decreases then weightage assigned to coverage gets increased and thus backbone nodes start going closer to host nodes in order to provide them more coverage as evident from figures $\ref{f5}~\text{and}~\ref{f6}$ for the value of $\lambda =0.5 ~\text{and}~0.1$ respectively.\\

\newpage

 The success of Steepest descent method depends on effective choice of both the descent direction and step size. In computing the step size $\delta_k$ there exists a tradeoff as we would like to select $\delta_k$ such that it gives a substantial reduction in $f$, but at the same time we do not want to give too much time making the choice but in general, it is too expensive to identify this value.\\

\begin{figure*}[htb!]
\begin{center}
\includegraphics[height=4cm, width=6cm]{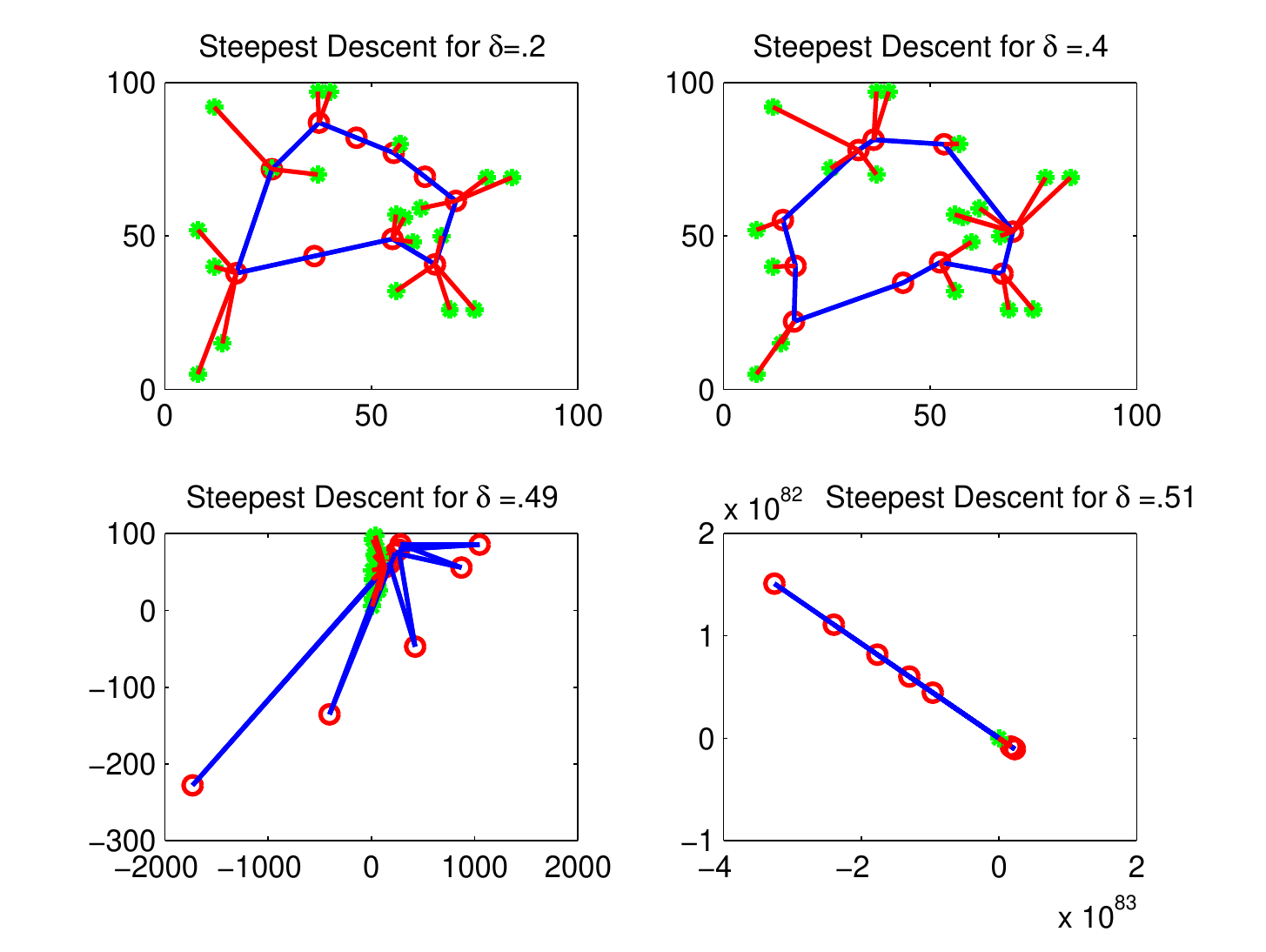}
\end{center}
\caption{Optimized network by Steepest Descent for $ \lambda =1$}\label{f7}
\end{figure*}

\begin{figure*}[htb!]
\begin{center}
\includegraphics[height=4cm, width=6cm]{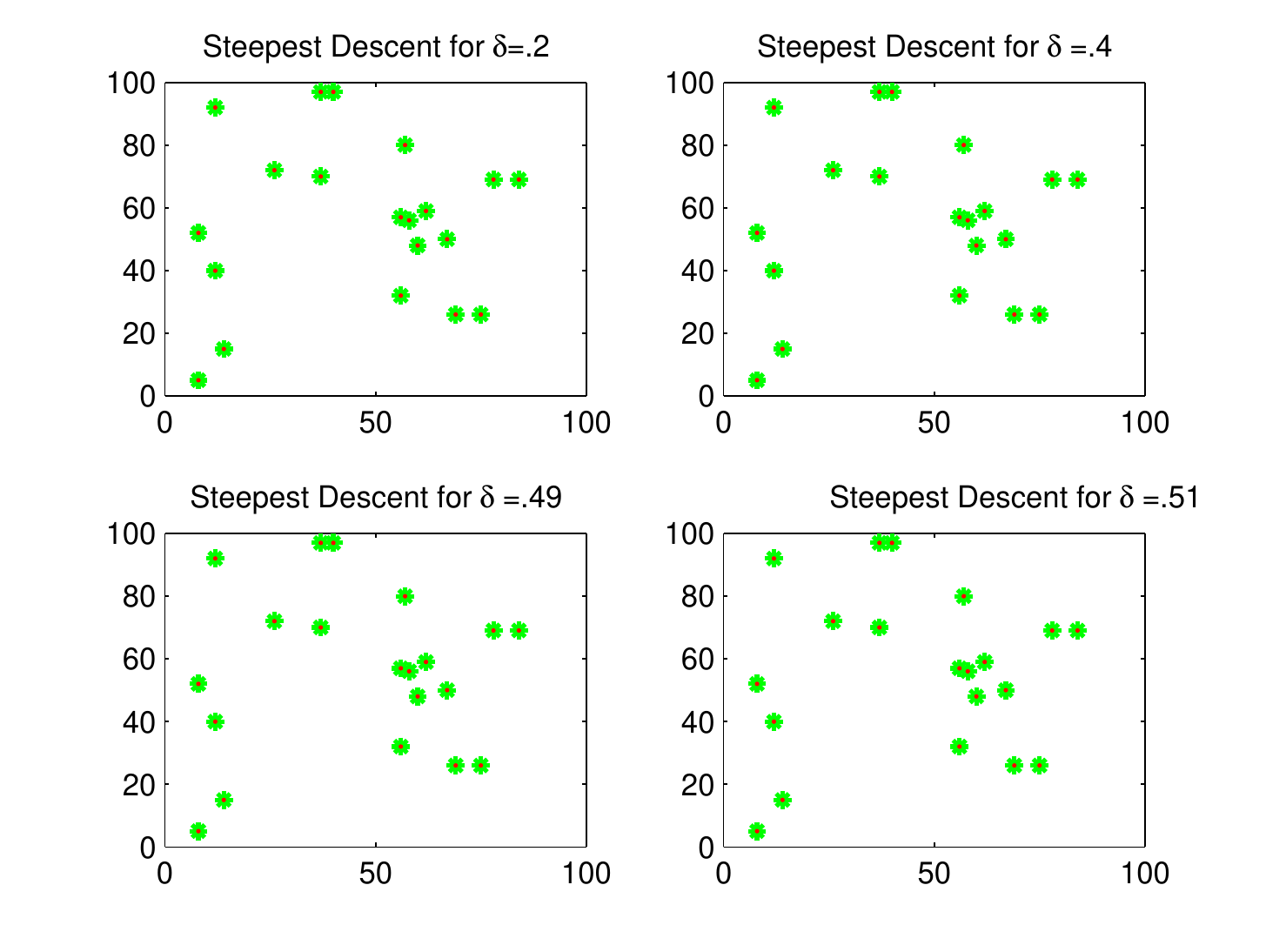}
\end{center}
\caption{Optimized network by Steepest Descent for $ \lambda =5 $}\label{f8}
\end{figure*}

\begin{figure*}[htb!]
\begin{center}
\includegraphics[height=4cm, width=6cm]{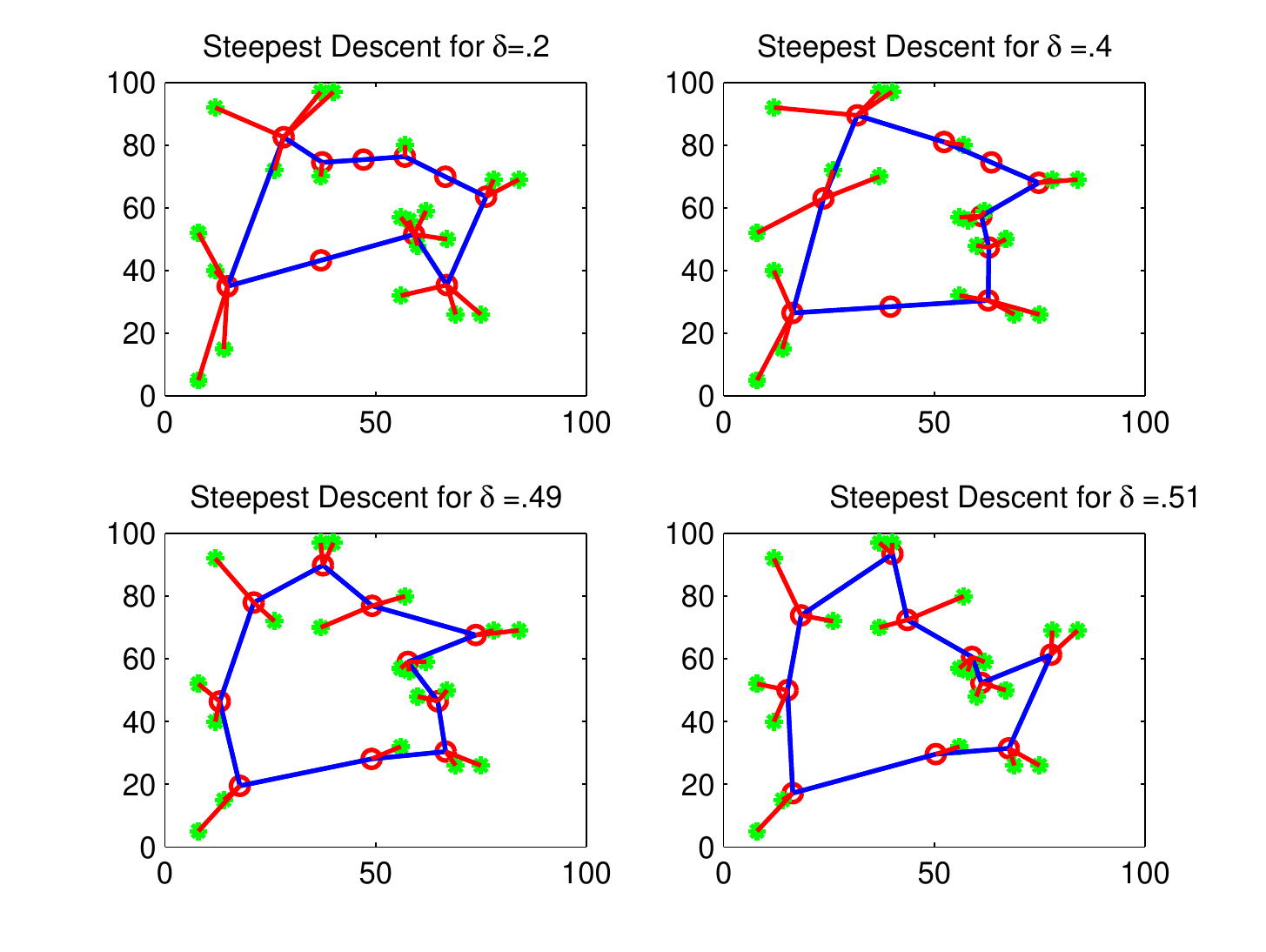}
\end{center}
\caption{Optimized network by Steepest Descent for $ \lambda =.5$}\label{f9}
\end{figure*}

For simulation purpose, we have assumed that for any randomly generated initial network, every backbone node is connected with two other backbone nodes (to maintain connectivity) and to atmost 20 host nodes (to maintain coverage). Based on this assumption and using (\ref{e31}), we find an interval containing desirable step size. Since sufficient step size $\delta_k$ is necessary in order to ensure better optimization as shown in figure $\ref{f7}$, $\ref{f8}$, and $\ref{f9}$. We choose manually some value of step size from obtained interval of desirable step size.\\

For $\lambda =1$ and the step size  $\delta \in (1/22,1/2)$, if the step size increases and move towards right extremity of this interval, we get networks in which backbone nodes starts relocating outside the assumed plane as shown in figure $\ref{f7}.$ For example, if the step size $\delta =.51 > .5$ (i.e. outside the interval) we don't get an optimized network and it also leads to increase in cost function.\\

For $\lambda =5$ and step size $\delta \in (1/30,1/10)$, no optimized network is obtained as all step size taken in figure are greater than right extremity i.e. $1/10.$ as shown in figure $\ref{f8}$\\

Similarly, for $\lambda =.55$ and step size $\delta \in (1/21,1),$ an optimized network is obtained since all step size lies within the interval as shown in figure $\ref{f9}$.\\

\newpage
For performance evaluation, algorithms are compared on the basis of optimized cost, number of iteration and total time taken to get an optimized network. Since algorithms are executed every time for different initial networks, therefore comparison has been made on the average taken from $20$ independent networks. For comparison, $\lambda=1$ has taken.\\

\begin{figure*}[htb!]
\begin{center}
\includegraphics[height=4cm, width=6cm]{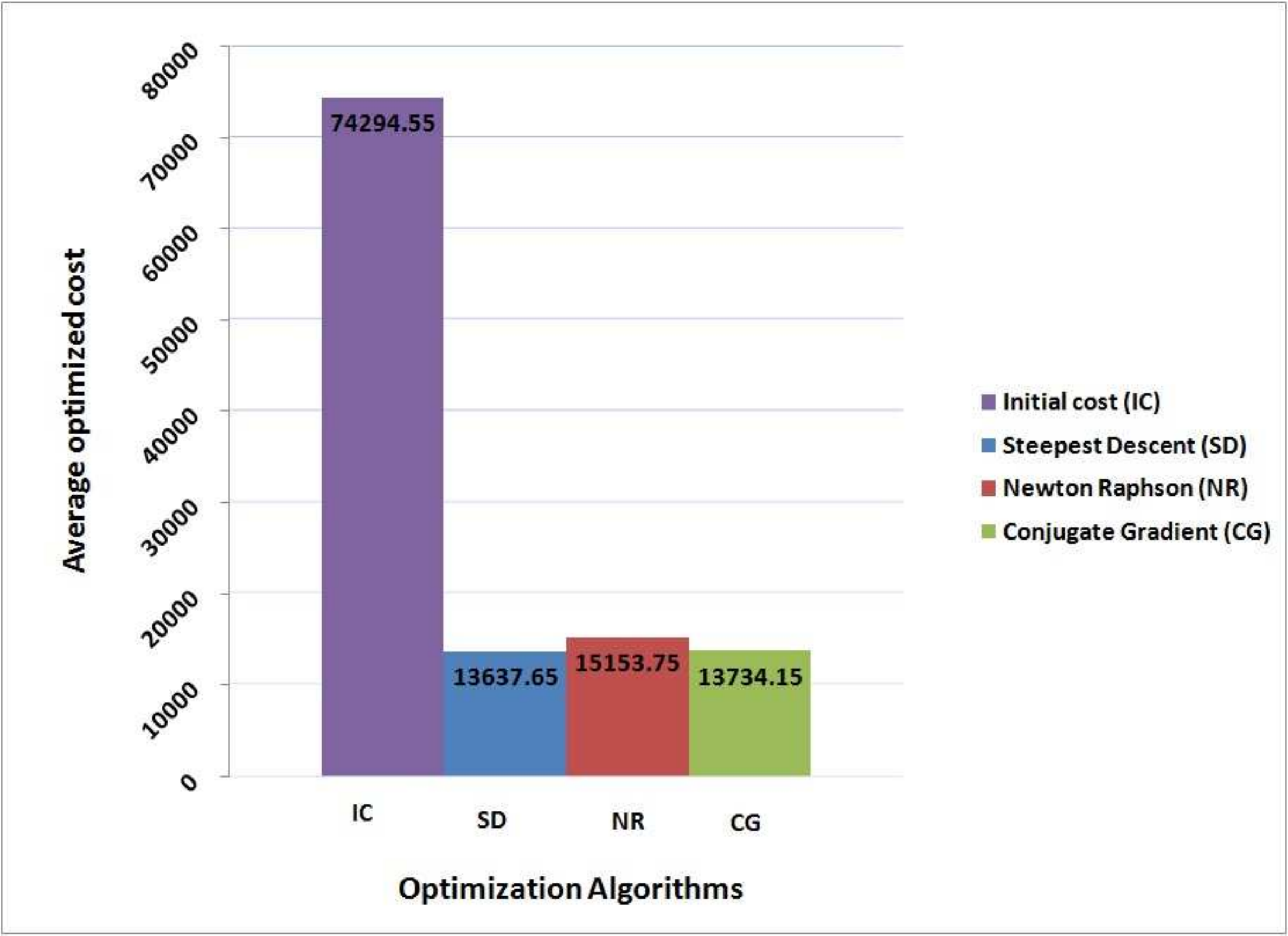}
\end{center}
\caption{Average optimized cost by different algorithms}\label{f10}
\end{figure*}

Figure $\ref{f10}$ presents the comparison of different optimization algorithms in terms of average optimized cost (average over $20$ independent networks). From this figure it can be seen that the initial cost of the network is approximately $74294$ and the cost of optimized network is in the range of $13637~to~15154$. Among the different optimization algorithm, the Steepest Descent gives minimum cost. It can be also observe from the figure that, Steepest Descent and Conjugate Gradient have almost similar performance in terms of average optimized cost.\\

\begin{figure*}[htb!]
\begin{center}
\includegraphics[height=4cm, width=6cm]{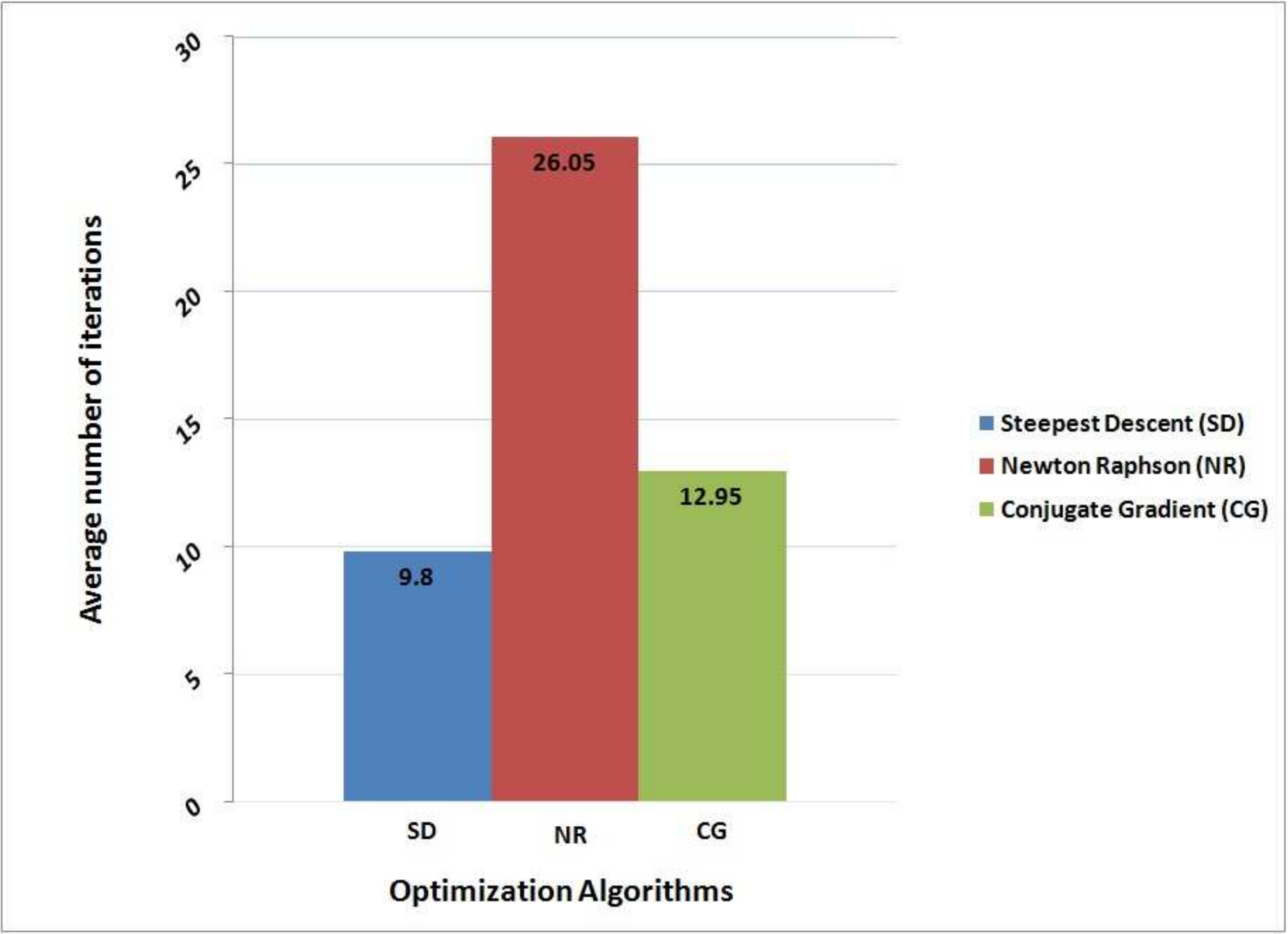}
\end{center}
\caption{Average number of iteration by different algorithms}\label{f11}
\end{figure*}

Figure $\ref{f11}$ shows the comparision of average number of iterations (average over $20$ independent networks) taken by each algorithm. From this figure it can be observed that, Steepest Descent method takes least number of iterations. Between Newton-Raphson and Conjugate Gradient method, Conjugate Gradient has relatively lesser number of iterations, therefore it may be concluded that Steepest Descent is best among all optimization methods in terms of number of iterations.\\

\begin{figure*}[htb!]
\begin{center}
\includegraphics[height=4cm, width=6cm]{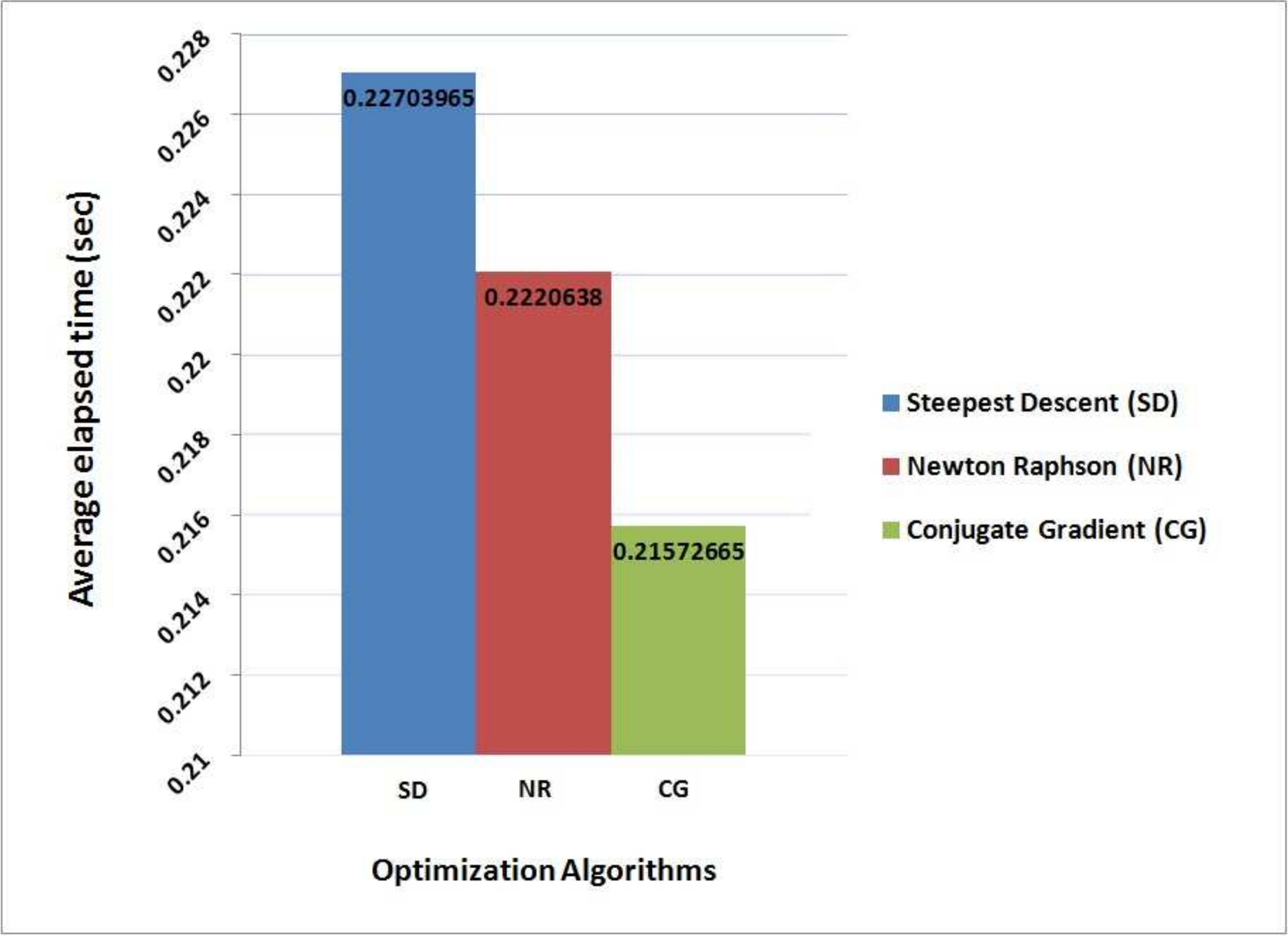}
\end{center}
\caption{Average time taken by different algorithms}\label{f12}
\end{figure*}

Although the Steepest Descent has lesser number of iterations but its elapsed time is relatively higher than Conjugate Gradient method as shown in figure $\ref{f12}$. It means that the Conjugate Gradient method takes lesser time to perform an iteration.\\

\newpage
\begin{figure*}[htb!]
\begin{center}
\includegraphics[height=4cm, width=6cm]{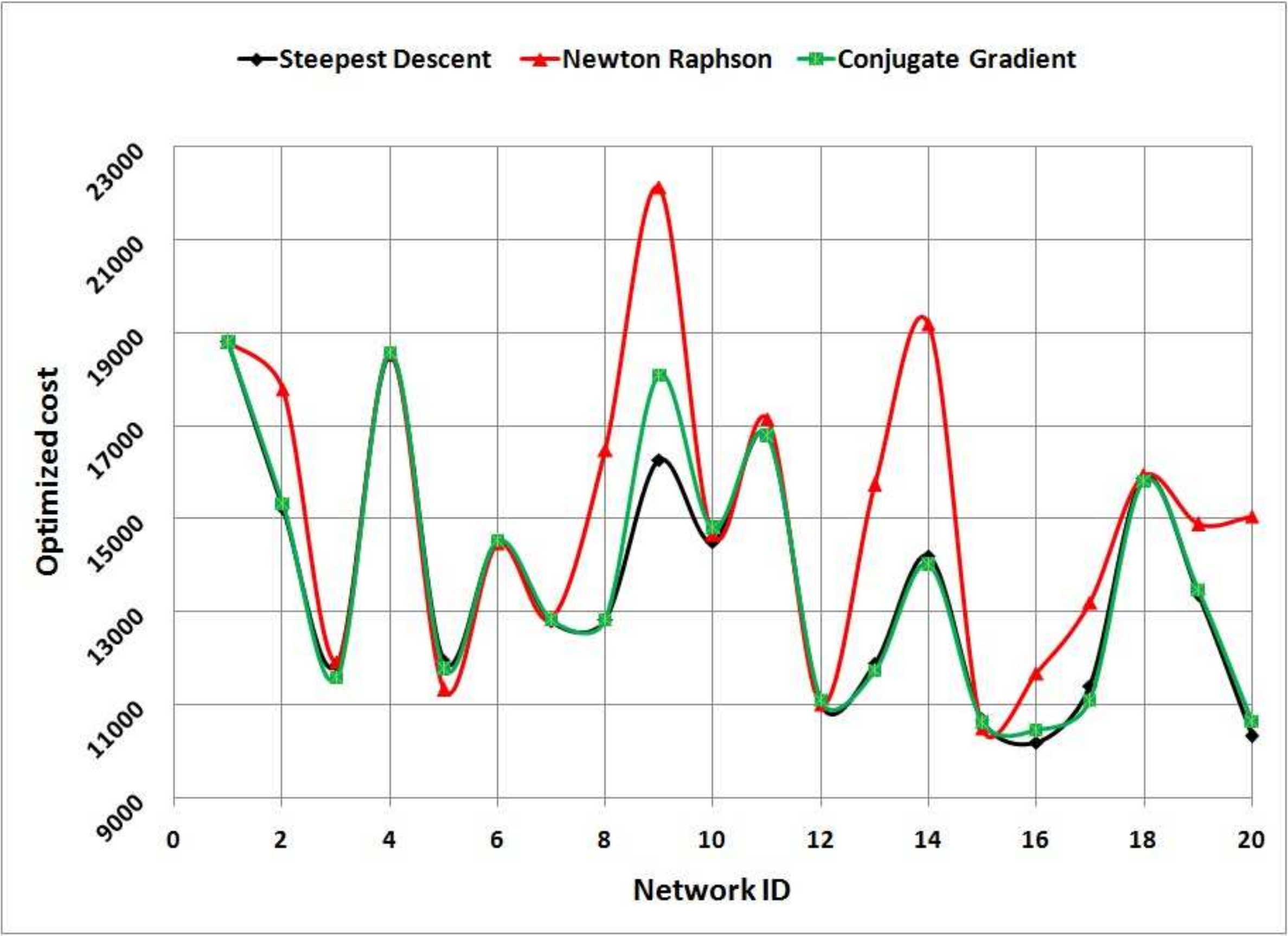}
\end{center}
\caption{`optimized cost' of 20 random network by different algorithms}\label{f13}
\end{figure*}

\begin{figure*}[htb!]
\begin{center}
\includegraphics[height=4cm, width=6cm]{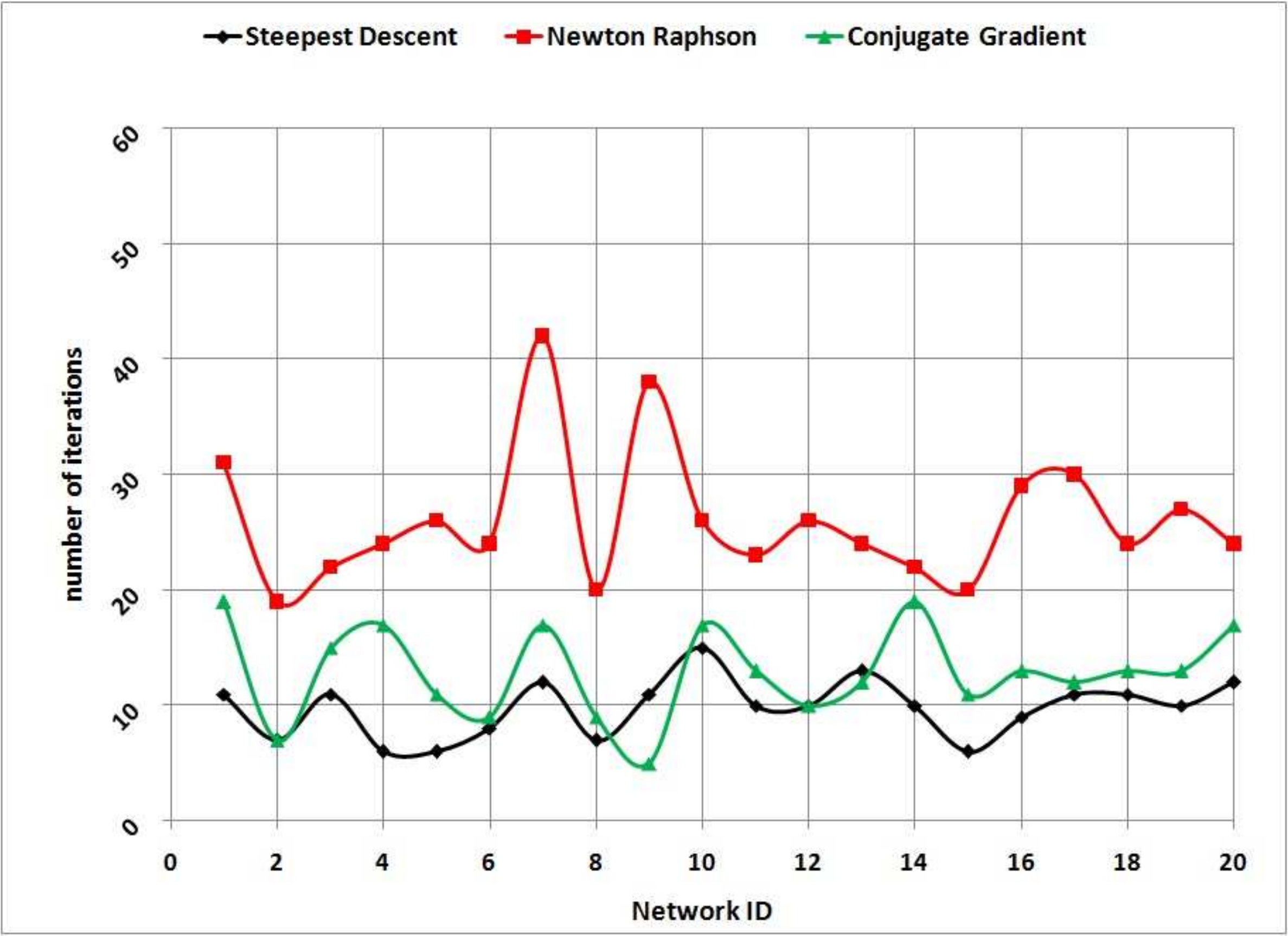}
\end{center}
\caption{`number of iteration' by different algorithms to optimized 20 random network }\label{f14}
\end{figure*}

\begin{figure*}[htb!]
\begin{center}
\includegraphics[height=4cm, width=6cm]{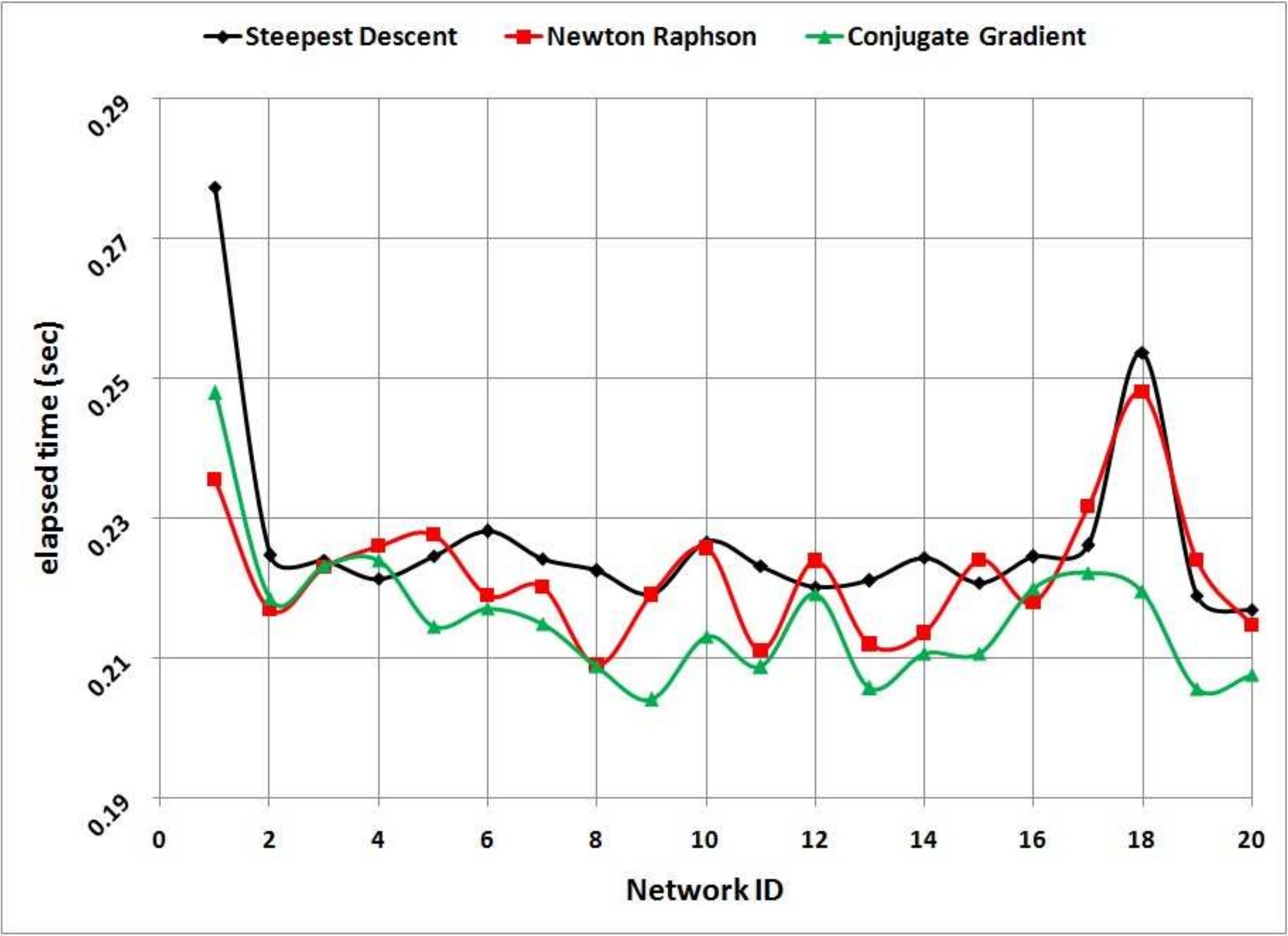}
\end{center}
\caption{`time taken' by different algorithms to optimized 20 random network}\label{f15}
\end{figure*}

Further in order to support the discussion made earlier on the figure $\ref{f10},\ref{f11}~ \text{and}~\ref{f12}$, the value of corresponding metrics for each of the $20$ independent networks are shown in figure $\ref{f13},\ref{f14}~ \text{and}~\ref{f15}$ to verify the consistency of results.\\


\newpage
\section{Conclusion}
In this paper, performance evaluation of Newton-Raphson and Conjugate Gradient method has been studied in comparison to Steepest Decent method for optimizing network coverage and backbone connectivity in backbone based wireless networks by adjusting the position of backbone nodes. Through simulation it has been found that Conjugate gradient method takes lesser time than other methods to optimize both coverage and connectivity but Steepest descent method does better optimization of coverage and connectivity. Thus there exists trade-off among the results obtained by Steepest Descent, Newton-Raphson and Conjugate Gradient method. When accuracy is needed, Steepest Descent is preferred, while if time is more important factor, Conjugate Gradient method is preferred.\\

    Since we have not taken into consideration the real world constraints such as power limitations, link blockage by terrain and capacity of base stations, therefore in our future work we will try to optimize the network coverage and backbone connectivity by using different optimization techniques and evaluate their performances by considering the above real world physical constraints.

\end{document}